\documentclass[twocolumn,trackchanges]{aastex701}
\usepackage{amsmath, amsthm}

\begin{document}

\title{Compton hump reverberation lag in the bright Seyfert\,1 galaxy IC\,4329A with {\it NuSTAR}}

\author[]{Samuzal Barua}
\affiliation{Shanghai Astronomical Observatory, Chinese Academy of Sciences, 80 Nandan Road, Shanghai 200030, China}
\email[]{samuzal.barua@gmail.com}
\correspondingauthor{Samuzal Barua}

\author[]{Hengxiao Guo} 
\affiliation{Shanghai Astronomical Observatory, Chinese Academy of Sciences, 80 Nandan Road, Shanghai 200030, China}
\email{}

\author[]{Minfeng Gu}
\affiliation{Shanghai Astronomical Observatory, Chinese Academy of Sciences, 80 Nandan Road, Shanghai 200030, China}
\email{}

\author{Wenwen Zuo}
\affiliation{Shanghai Astronomical Observatory, Chinese Academy of Sciences, 80 Nandan Road, Shanghai 200030, China}
\email{}

\begin{abstract}

Recent reverberation delay measurements have moved beyond the 10 keV X-ray range, providing evidence for the Compton hump (a.k.a. reflection hump) in the lag spectra. We report the relativistic reverberation of the reflection hump in the bright Seyfert\,1 galaxy IC\,4329A based on a long {\it NuSTAR} observation. We find a delayed response of the 20--30 keV X-ray band, with a lag time of $\sim1825$ s at frequencies $0.5-1.5 \times 10^{-4}$ Hz. The lag amplitude drops to $\sim195$ s as the frequencies increase to $(1.5-10)\times10^{-4}$ Hz. Including IC\,4329A, so far five sources have been explored for reflection hump reverberation. We perform reverberation modelling of the 3--50 keV lag-energy spectra using the general relativistic transfer function code, which provides independent timing-based measurements of the black hole mass $M_{\rm BH}=1.37_{-0.36}^{+0.33}\times10^8~M_{\odot}$ and the coronal height $h=2.45_{-2.36}^{+1.92}~R_{\rm g}$ (with uncertainties at 90\% confidence). Within the uncertainties, the measured mass is found to be consistent with the previous finding. Furthermore, we undertake reflection spectroscopy to account for the hump feature and the associated relativistic effect using the time-averaged flux spectrum. Further sampling of the {\it NuSTAR} data (with a bin width of 0.2/0.4 keV below and above 10 keV) that reshapes the spectral resolution allows us to constrain the coronal temperature at $50.26_{-4.03}^{+5.58}$ keV -- consistent with the previous result from the combined {\it Suzaku} and {\it NuSTAR} data.
\end{abstract}
\keywords{\uat{Black hole physics}{159} --- \uat{Accretion}{14} --- \uat{X-ray active galactic nuclei}{203} --- \uat{Reverberation mapping}{2019} --- \uat{Seyfert galaxies}{1447}}


\section{Introduction} 
X-ray reflection spectrum produced in the vicinity of the black hole becomes highly distorted by the combined effect of Doppler boosting and gravitational redshift. 
Commonly appearing reflection features that are imprinted on the observed spectrum generally include the soft X-ray excess below 1 keV and the Fe K$\alpha$ line at 6.4--6.97 keV (depending on the ionization state) and the reflection hump at $\sim20-30$ keV \citep{George1991, Turner1999}.  
Under the relativistic effects, the observed features appear relativistically blurred, in particular, resulting in a broadened and skewed iron line profile \citep{Fabian1989}. 

Over the past decade or so observations have unveiled the relativistic reverberation response of the inner accreting region around the black hole, in which a time delay of the disc-reflected photons appears with respect to the power law continuum from the corona \citep{Uttley2014}. The reverberating signal has been attributed to the soft excess, first observed in the Narrow-line Seyfert\,1 (NLS1) galaxy 1H0707-495 \citep{Fabian2009Natur}, and several other targets in successive campaigns \citep[e.g.][]{Emmanoulopoulos2011, DeMarco2011, Cackett2013, Zoghbi2011REJ1034+396, Wilkins2017, Alston2020}. Objects in considerable quantity have also shown that the tracing of the Fe K$\alpha$ line is the aspect of relativistic reverberation from the innermost region of the disc \citep[e.g.][]{Zoghbi2012NGC4151, Zoghbi2013FeKlag, Marinucci2014, Kara2014, Alston2015, Kara2016FeKlag, Wilkins2021Natur} -- the illumination pattern of which is imprinted on the line profile itself. It has been shown that Fe K$\alpha$ lag amplitude can be used as an indicator of the black hole mass \citep[see][]{Uttley2014, Kara2016FeKlag}.


Sources exhibiting strong X-ray reflections, being that the pronounced reflection hump would typically appear with a relativistic reverberation delay. This has been possible with the {\it Nuclear Spectroscopic Telescope Array} ({\it NuSTAR}) \citep{Harrison2013}, the hard X-ray focusing telescope in orbit that extends the sensitivity and focusing optics far beyond $\sim10$ keV. So far, search for reflection hump reverberation has been explored in the Seyfert 1 galaxies SWIFT J2127.4+5654, NGC 1365 \citep{Kara2015NGC1365}; MCG-5-23-16 \citep{Zoghbi2021} and Ark 564 \citep{Lewin2022}, which show characteristic hump in their flux spectra. The spherical lamppost geometry has also been tested in the three sources, SWIFT J2127.4+5654, MCG-5-23-16 and Ark 564 by \citep{Zoghbi2021, Lewin2022} through reverberation modelling of their lag spectra.
Like lag measurements in the soft X-ray and Fe K emission bands in some other sources, hard lag has not been reported in any of these sources with an absolute lag value in the $\sim20-30$ keV band with respect to the below $\sim20$ keV or above the $\sim40$ keV. However, the underlying hump feature has been found to contribute to the above 10 keV lags.

Reflection hump can be used as an ingredient to understand the coronal properties; in terms of electron temperature ($kT_{\rm e}$), optical depth ($\tau$), and compactness parameter ($l$) \citep{Fabian2015}. {\it NuSTAR} has added a new dimension to reflection spectroscopy, which enables constraining the hump-like feature in the flux spectrum, which in turn determines the coronal temperature via the high-energy cutoff -- the exponential rollover energy $E_{\rm cut}=2-3~kT_{\rm e}$ \citep{Zdziarski1996, Zycki1999}. Analysis of the prominent reflection hump in the {\it NuSTAR} spectrum yielded a coolest temperature corona ($kT_{\rm e}=15\pm2$ keV) in the X-ray bright NLS1 Ark 564 from reflection spectroscopy \citep{Kara2017}. 
%
While relativistic blurring effect is seen in the flux spectrum, in particular, in the energy range where reflection hump peaks \citep{Reynolds2021}, the shape of the lag spectrum reproduced should typically demonstrate the same at high frequencies and can be described by reverberation modelling for a high spin case. 

One of the good candidates in which the reverberation delay of the reflection hump can be observed is the one showing that feature in the flux spectrum. The {\it NuSTAR} spectrum of the bright Seyfert 1 galaxy IC\,4329A (z=0.0161) revealed the presence of the clear reflection hump \citep{Brenneman2014, Tortosa2024}, which has placed this target at the forefront of the search for reverberation delay beyond 10 keV X-ray. The target with a black hole mass of $M_{\rm BH}=1.2\times10^8~M_\odot$ \citep{deLaCalle2010} presents itself to be brightest in the X-ray sky, with a 2--10 keV flux $F_{2-10~\rm keV}$ of $(0.1-1.8)\times10^{10}~\rm erg~cm^{-2}~s^{-1}$ \citep{Beckmann2006}. Its coronal properties have been explored in a number of studies, providing high-energy cutoff $E_{\rm cut}$ measurements $\sim 150-390$ keV \citep{Gondoin2001, Molina2009, Brenneman2014, Ricci2017, Pal2024, Tortosa2024}. Based on spherical geometry, the coronal temperature $kT_{\rm e}$ has been found to be at $50_{-3}^{+6}$ keV and the optical depth $\tau=2.34_{-0.11}^{+0.16}$ \citep{Brenneman2014}. 

Having observed the usual appearance of the reflection hump in the X-ray spectrum of IC\,4329A, here we extend the study to search for relativistic reverberation delay in the hard X-ray band using {\it NuSTAR} observation. We also carry out spectral analysis to constrain the coronal temperature via the reflection hump.
The paper is organized as follows: In Section 2, we present the {\it NuSTAR} observation and data reduction, and in Section 3 we present briefly the Fourier-based timing analysis. We present the time lags and the results from reverberation and spectral modelling in Section 4. In Section 5, spectral timing results are discussed and interpreted.

\section{Observation and data reduction} \label{sec:style}

IC\,4329A was observed by {\it NuSTAR} for $\sim162$ ks exposure (Obs. ID 60001045002) in 2012, which is the longest observation obtained for this source with {\it NuSTAR}.
We reduce the data by running the standard pipeline {\sc nupipeline} which is part of the {\sc heasoft} package. This provides Level-1 data products. The data were screened for background flare and the standard filtering criteria was applied. {\it NuSTAR} data analysis software {\sc nustardas} (v.1.10) was used with the calibration database ({\sc caldb}; v.20250415) to obtain cleaned and calibrated Level-2 event files. We used a circular extraction region of 50\arcsec  ~in radius for the source and a larger background region with 90\arcsec ~in radius for modules A and B. With the {\sc nuproducts} task we produce the final data products, which generally include the spectra and light curves. For timing analysis, we produce light curves in 11 narrow energy bins across the 3--50 keV band, each with a time bin size of 600 s. The spectra are grouped using the standard procedure (presented in Section 4.2). 

\begin{figure}
\vspace{-2.5cm}
\hspace{-1.2cm}
\includegraphics[scale=0.46, angle=-90]{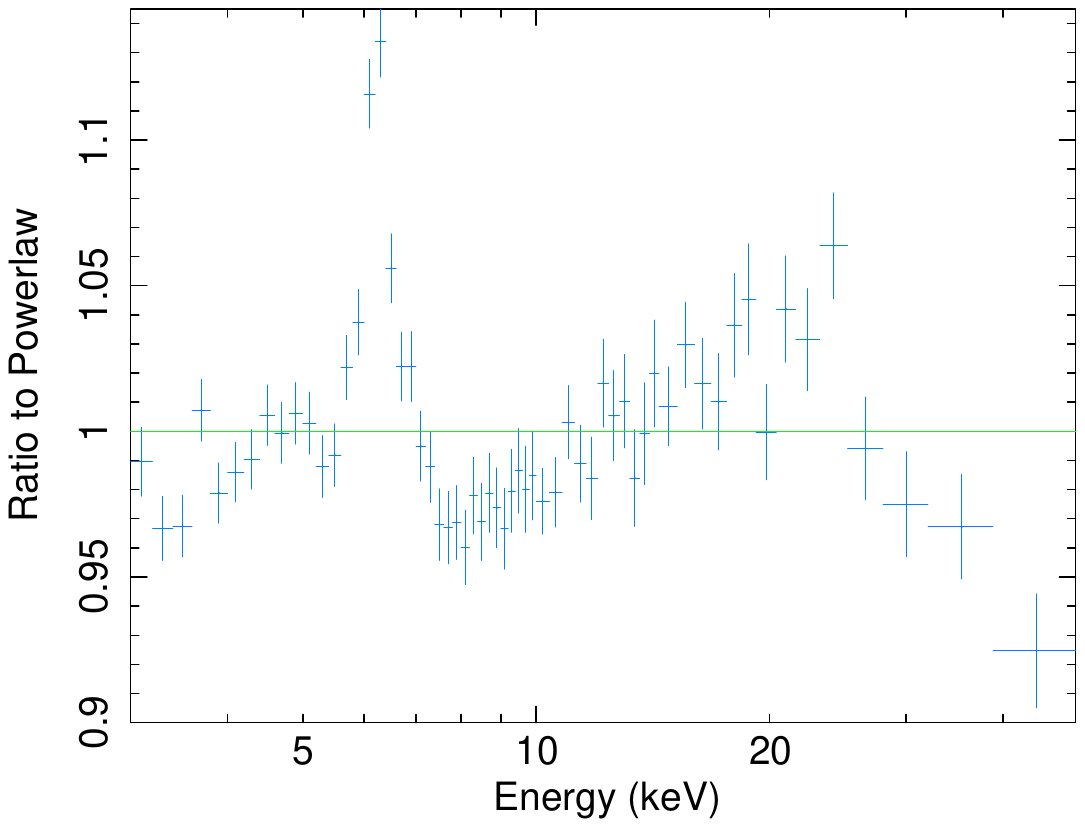}
\vspace{-0.8cm}
\caption{Zooming the ratio of the {\it NuSTAR} spectrum to the simple power law model. The plot shows the presence of a strong narrow Fe K emission line at $\sim6.4$ keV and a clear Compton hump feature at $\sim$20--30 keV. The data is rebinned for visual clarity.
\label{fig:ratio_plot}}
\end{figure}

\begin{figure*}
\includegraphics[scale=0.45, angle=0]{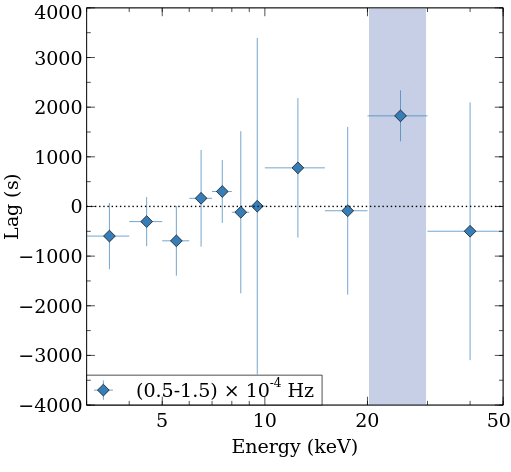}
\includegraphics[scale=0.45, angle=0]{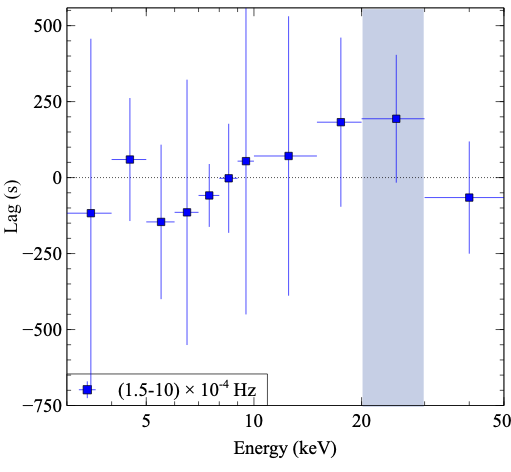}
\includegraphics[scale=0.45, angle=0]{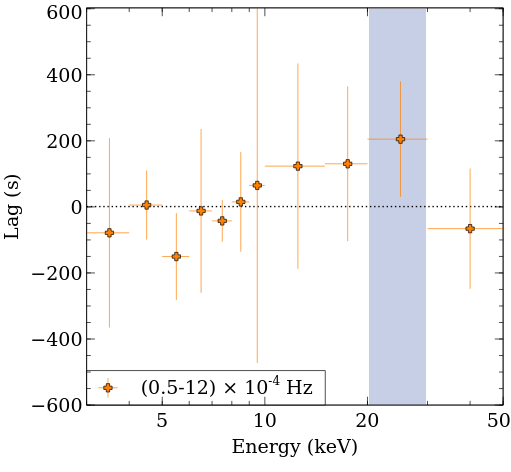}
\caption{High frequency lag-energy spectra of IC\,4329A. From left to right the lag spectra are calculated for frequencies $(0.5-1.5)\times10^{-4}$ Hz, $(1.5-10)\times10^{-4}$) Hz, and $(0.5-12)\times10^{-4}$) Hz, respectively. The plots show the clear delayed response of the reflection hump at 20--30 keV band (indicated by the vertical shaded region), with lag amplitudes of $\sim1825$ s, $\sim195$ s, and $\sim205$ s from left to right, respectively. A broader frequency range is used in the right panel to increase the signal-to-noise.}
\label{high-freq_lag_spec}
\end{figure*}

\section{Fourier based time lags with Gaussian Process}
Calculation of the time lags between light curves generally requires evenly sampled X-ray counts. The direct Fourier method has the potential to provide robust lag measurements, which have been obtained mostly from the {\it XMM-Newton} observations due to its large effective area, which provides a sufficient count rate. The standard Fourier method for calculating X-ray time lags was presented by \citet{Nowak1999} and was further outlined by \citet{Uttley2014}. This approach generally cannot be applied to the {\it NuSTAR} data, as the data obtained have gaps of 90 minutes due to Earth occultation as the observatory passes through the South Atlantic Anomaly (SAA). Therefore, to deal with irregularly spaced data, another method, the Gaussian Process Regression (GPR), has been developed. We use GPR on the {\it NuSTAR} data followed by the standard Fourier method on the interpolated light curves to compute the lag spectra. For a detailed introduction of GPR, we refer to \citet{Rasmussen2004, Wilkins2019, Griffiths2021}. However, one can alternately use the Maximum likelihood approach \citep{Miller2010, Zoghbi2013} to calculate time lags by interpolating the gaps in the light curve (Barua et al. {\it in preparation}). Given the observed data, the likelihood method relies on the most likely variability power and time lag. A model is first constructed in the functional form of power law and lag, where the model is parameterized by the values of power and lags in the pre-defined frequency bins. A likelihood function is constructed by comparing the model to the autocorrelation and cross-correlation of the data.  The best estimates are typically obtained by maximizing the likelihood function. The autocorrelation here is the Fourier transform of the PSD (power spectral density), and the cross-correlation is the Fourier transform of the cross-spectrum. Similar to the standard Fourier technique, the cross-spectrum provides the phase difference between the two light curves, which is converted into time lag by dividing by $2\pi f$ \citep{Zoghbi2013}. 

In a short recap, GPR is a non-parametric Bayesian method that assumes the underlying function (source count rates as a function of time) is a realization of a Gaussian process. This is characterized by a mean and a covariance function. The mean function is generally set to zero, but can be customized depending on the data properties, while it represents the value of the function that is modeled at each input point. The covariance or kernel function encodes the assumption about the data's behaviour such as periodicity or smoothness, and determines the similarity of the data points in the light curve to each other. To ascertain the correlation between the values of the function, the kernel determines pairwise similarities. More typically, the kernel function models the covariance between two time points in different light curves as a function of their difference and the inherent lag. The often used Radial Basis Function (RBF) or Squared Exponential Kernel function is defined as: 
\begin{equation}
    k(x_i,x_j)= \sigma^2 exp\Biggl(-\frac{||x_i -x_j||^2}{2l^2}\Biggl)
\end{equation}
where, $\sigma^2$ is the variance of the kernel function that determines how strongly the data points are correlated. A higher value of the variance is represented by the higher $\sigma^2$. $|x_i -x_j||^2$ represents the geometric separation of the distance between the data points (known as the squared Euclidean distance) and $l$ the characteristic length, which represents that as the data pints are spaced apart the kernel function degrades. The selected kernel has hyperparameters such as the length scale, signal variance, and noise variance, which are optimized (i.e. the GPR model is trained on the light curves) by maximizing marginal likelihood of the observed data. The optimized kernel (or the trained GPR model) effectively interpolates the gaps, producing a complete, contentious light curve for the given energy bands by calculating the posterior distribution, where the best prediction is obtained from the mean of the distribution and the uncertainty obtained from the variance.
Thus, the trained GPR model can be used to calculate the time lags by employing the standard Fourier method as follows: 

The Fourier transforms of the light curves are taken, say the band of interest $x(t)$, and the reference band $y(t)$. The Fourier-transformed light curves $\tilde X(\nu),\, \tilde Y(\nu)$ are multiplied, one by the complex conjugate of the other, which constructs a cross-spectrum $C(\nu)$ defined as
\begin{equation}
\begin{split}
    C(\nu) = \tilde X(\nu) \tilde Y^*(\nu)=|\tilde X(\nu)|e^{i\phi_x}~|\tilde Y(\nu)|e^{-i\phi_y}\\
    =|\tilde X(\nu)|~|\tilde Y(\nu)|~e^{i(\phi_x - \phi_y)}
\end{split}   
\end{equation}

The cross spectrum encodes the phase phase difference between the two light curves which is transformed into the time lag $\tau(\nu)$ by dividing by $2\pi\nu$, producing frequency-dependent lags
\begin{equation}
    \tau(\nu)= \frac{\rm arg[C(\nu)]}{2\pi\nu}
\end{equation}

Here we create the lag energy spectrum by constructing the cross-spectrum between 1000 of GP realizations of each band of interest and the reference band. The cross-spectrum is averaged over a broad frequency bins to optimize the signal-to-noise. A broad reference band is taken, covering the entire energy range 3--50 keV, but not including the current bands of interest.  The use of a broad reference band is to reduce the correlated noise between the narrow bands. We use the GPR implementation of the Fourier method using the package pyLag\footnote{https://github.com/wilkinsdr/pyLag} \citep{Wilkins2019} and the standard library {\tt Scikit-learn}\footnote{https://scikit-learn.org/} for the basic GPR task.

\subsection{Choice of the Gaussian Process Kernel}
 Implementation of the Gaussian Process (GP) kernels becomes successful in recovering Fourier lags in the data from the telescopes on the low-Earth orbit, such as {\it NuSTAR}, {\it NICER} and {\it STROBE-X} \citep{Wilkins2019}. It has been found that the top two kernels, Rational quadratic (RQ) or Matérn kernels, can provide accurate Gaussian process models of the observed light curves from these three telescopes \citep{Wilkins2019}. While the lag measurement was seen to relies on the choice of the GP kernels in the {\it Swift} data \citep{Griffiths2021}, their impact, on the other hand, is much less apparent in the {\it NuSTAR} data \citep{Wilkins2019, Lewin2022}. This has been accounted for, likely due to the difference in the data sampling/cadence and the length between the {\it Swift} and {\it NuSTAR} observations. 

 We present the common functional forms of the above two GP kernels. While a single length scale parameter, $l$, in the RBF kernel is unable to account for the time scale variability of the light curve, it has been found appropriate to select sum squared exponential functions
with different length scales \citep{Wilkins2019}. An infinite sum of series of squared exponential functions with different length scale parameter provides the RQ kernel, which can be expressed as
\begin{equation}
    k(x_i,x_j)= \Biggl(1+ \frac{d(x_i,x_j)^2}{2\alpha l^2}\Biggl)^{-\alpha}
\end{equation}
Where $\alpha$ is the scale mixture parameter, which changes the value of the length scale parameter $l$. The RQ kernel is generally seen as the scale mixture of the RBF  kernel with different $l$ values. The other form of GP kernels, Matérn kernel, is also a generalized form of the RBF kernel. In addition to the length scale parameter $l$, the kernel includes another parameter $\nu$ to control the smoothness of the function. The resulting function is defined as
\begin{equation}
   k(x_i,x_j)= \frac{1}{\Gamma( \nu)2^{\nu -1}} \Biggl(\frac{\sqrt{2\nu}}{l}d(x_i,x_j)\Biggl)^\nu K_\nu \Biggl(\frac{\sqrt{2\nu}}{l}d(x_i,x_j)\Biggl) 
\end{equation}
Where $K_\nu$ is a modified Bessel function and $\Gamma$, the gamma function. As $\nu=1/2$ the Matérn kernel transforms to an absolute exponential kernel. When $\nu \to \infty$ the Matérn kernel converges to RBF kernel.

We use both the GP kernels, RQ and Matérn kernels, and observe that the choice of their forms has no significant impact on the Fourier lag detections, providing stable lag measurements. Our findings appear to be consistent with those of \citet{Lewin2022} who similarly recovered lags using the {\it NuSTAR} data. We notice that the uncertainties on the lag values are more or less similar regardless of the choice of kernels. The uncertainties appear within about 10\% on-average across three different sets of frequencies (see the next section), while the lag values appear within 10\% at low frequencies and 20\% at high frequencies. We note, however, that the RQ kernel provides a marginally better description of light curve variability with a lower negative log marginal likelihood (NLML of 67.05) than the Matérn kernel, which provides NLML of 67.5. Using {\it NuSTAR} observation of Ark\,564 \citet{Lewin2022} also observed that the RQ kernel is most useful for recovering lags. This was earlier inferred by \citet{Wilkins2019} based on {\it NICER} observation of the same source. The authors further noted that sufficient accuracy can be achieved with the use of the RQ kernel, which in turn successfully recovers lags with on-orbit missions in low-Earth orbit. 

\begin{figure}
\includegraphics[scale=0.6, angle=0]{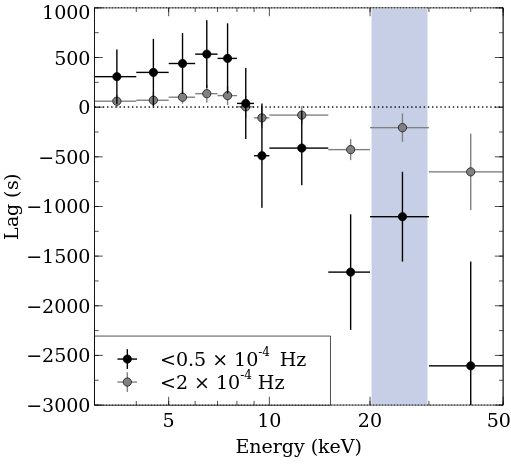}
\caption{Low-frequency lag-energy spectra of IC\,4329A, computed for frequencies below $0.5\times10^{-4}$ Hz and $2\times10^{-4}$ Hz. Two frequencies are shown for demonstrative purposes, indicating that the lag decreases as the frequencies are increased. In each spectrum, there appears a peak in the 20--30 keV band where the reflection hump dominates.}
\label{lowfreq_spec}
\end{figure}

\section{Results}

\subsection{Energy-dependent lags}

The ratio of the {\it NuSTAR} data to a simple power law model for IC\,4329A is produced in Figure\,\ref{fig:ratio_plot}, which shows the presence of the reflection hump at an X-ray energy above $\sim20$ keV.
Therefore, in order to search for its relativistic reverberation origin 
we compute lag-energy spectra across the 3--50 keV of the {\it NuSTAR} band, which represents lag in each narrow energy band of interest, measured relative to the reference band. Positive lags on the Y-axis refers to the band of interest lags with respect to the reference band. 

Figure\,\ref{high-freq_lag_spec} shows the lag-energy spectra of IC\,4329A calculated for different sets of high frequencies. Calculating the energy-dependent lags at frequencies $(0.5-1.5)\times10^{-4}$ Hz, we see that the lag spectrum traces a peak lag in the 20--30 keV band where the reflection hump dominates. The delayed response of this reflection-dominated {\bf band} is seen with a lag amplitude of $\sim1825$ s with respect to the power law-dominated bands 8--9 keV or 15--20 keV ({\it left panel}). The observed lag amplitude drops as the frequencies are increased, but without changing much in the shape of the overall spectrum -- the whole lag spectrum shifts downward with decreasing lag times at higher frequencies. We observe that the 20--30 keV lag amplitude decreases to $\sim195$ s as the frequencies are further increased to $(1.5-10)\times10^{-4}$ Hz ({\it middle panel}). 
To further trace the 20--30 keV lag and that it is dominated by high frequencies, we compute the lag-energy spectrum from a cross-spectrum, averaging over a frequency range from 0.5 to $12\times10^{-4}$ Hz. A broader frequency range is taken to overcome the low signal-to-noise, but the lag spectrum is not seen to be dominated by low frequency. This results in a clear hump feature in the lag spectrum with a 20--30 keV lag amplitude of $\sim205$ s ({\it right panel}). Further extending the lower limiting frequency to $0.5\times10^{-6}$ Hz we see that the lag spectrum does not change its shape where the signal-to-noise ratio is improved, providing a 20--30 keV lag of $\sim115$ s (Figure~\ref{low-high-freq_lag_spec}, shown for demonstrative purpose). Although the magnitude appears smaller, it is within the uncertainties on the lags shown in the middle and right panels of Figure~\ref{high-freq_lag_spec}. Unless the higher frequency limit is changed, the reflection hump feature remains unaltered, implying that a change in the high frequency determines the delay of the corresponding hard band. If we go beyond the frequency range of $\sim12\times 10^{-4}$, the lag amplitude drops, but the lag detection is not statistically significant due to Poisson noise.

We perform a statistical test to determine the significance level of the detected lags in the 20-30 keV band. X-ray reverberation feature can generally be seen in the lag-energy spectrum when the reflection-dominated band differs considerably from the continuum-dominated band. Therefore, we first perform a null hypothesis test using a simple log-linear model of the form $y=a+b\,{\rm log(x)}$ \citep[see][for example]{Kara2016FeKlag}. We first fit this model to the 8--10 keV band, which is dominated by powerlaw continuum, and extrapolate it to 10--50 keV band. The above 10 keV band is dominated by reflection, which comprises the hump-like feature. Given that lag in the continuum-dominated band generally appears to be zero, the null hypothesis in this case is assumed to be a flat line with a slope of zero. We calculate $\chi^2$ between the null hypothesis and the 10--50 keV band. The significance test based on $\chi^2$ distribution provides a confidence level of $>95$\% at which 20--30 keV lag of $\sim1825$ ks is detected. We also carry out Monte Carlo simulations to asses the statistical test. For that we produce 1000 simulated light curve pairs using the method outlined in \citet{Timmer1995}, which are based on observed PSD (power spectral density) \citep[e.g.][]{Emmanoulopoulos2013}. The simulated light curves are then scaled to the mean and standard deviation of the observed light curves. The Monte Carlo simulation shows that the hard lags appear above the null hypothesis at $>95$\% confidence. 

Computing lag-energy spectra at low frequency intervals, on the other hand, shows a significant change in the shape of the lag spectra, particularly at energies above 10 keV. We show the lag-energy spectra for low frequencies in Figure~\ref{lowfreq_spec}. Two sets of frequencies are used just to show that the lags decrease as the frequency increases. The lag amplitudes are observed to decrease with increasing energy, where the lag values appear negative. We see that the peak at the reflection hump peaking band 20--30 keV appears to be prominent, responding later than the 15--20 keV and 30--50 keV bands. 
This low-frequency behaviour of the lag spectra is not unusual, as has been observed in the Seyfert\,1 galaxy NGC 1365 \citep[][see Figure\,7]{Kara2015NGC1365}. The observed energy-dependent lags for this source show large error bars across 10--50 keV but the reverberating hump remains apparent.
Similar trends of the spectra with negative lags have also been seen in the Seyfert 1 galaxies Ark 564 \citep[][see Figure\,10]{Chainakun2016} and IRAS 13224-3809 by \citep[][see Figure\,5]{Kara2013IRAS}, in which reverberating signatures can be distinguished. While low-frequency hard lag has not been well understood, the currently accepted scenario is that lags are attributed to fluctuations in the accretion rate that propagate over the disc and are transferred to the corona \citep{Kotov2001, Arevalo2006}. For a black hole mass of $1.2\times10^8~M_{\odot}$ in IC\,4329A, the low frequencies signals below $<0.5/2\times10^4$ Hz emerge from up to a few hundreds  $R_{\rm g}$, while above this range, up to maximum of $12\times10^4$ Hz, reverberation origin is typical that corresponds to a short time-scale variability arising from a few tens of $R_{\rm g}$ of the black hole or even less. The source has also been found to have an ultrafast outflow (UFO) and a cloud of warm absorber \citep{Tortosa2024}, and it is also likely that their movement from our line of sight during observation dominates the low frequency lags \citep{Kara2015NGC1365, Wilkins2023}.  

We observe that there does not appear to be a significant peak in the 6--7 keV band at high frequencies (Figure\,\ref{high-freq_lag_spec}),
as lag detections are hampered by low variability power. However, lowering the Fourier frequencies to below $2\times10^{-4}$ Hz, the lag spectrum is seen to indicate the Fe K emission feature at $\sim6.4$ keV (Figure\,\ref{lowfreq_spec}). Although the peak observed to be not strong enough, its appearance cannot be ruled out, as the broadened line feature has been found in the energy spectrum \citep{Brenneman2014}.  
Note further that the shown lag spectra are from module A. The signal-to-noise in the data from module B is not good enough, and the lag detection is hindered by the low variability power, which if combined with that from module A degrades the quality of the spectra. 


\begin{figure}
\hspace{-0.5cm}
\includegraphics[scale=0.62, angle=0]{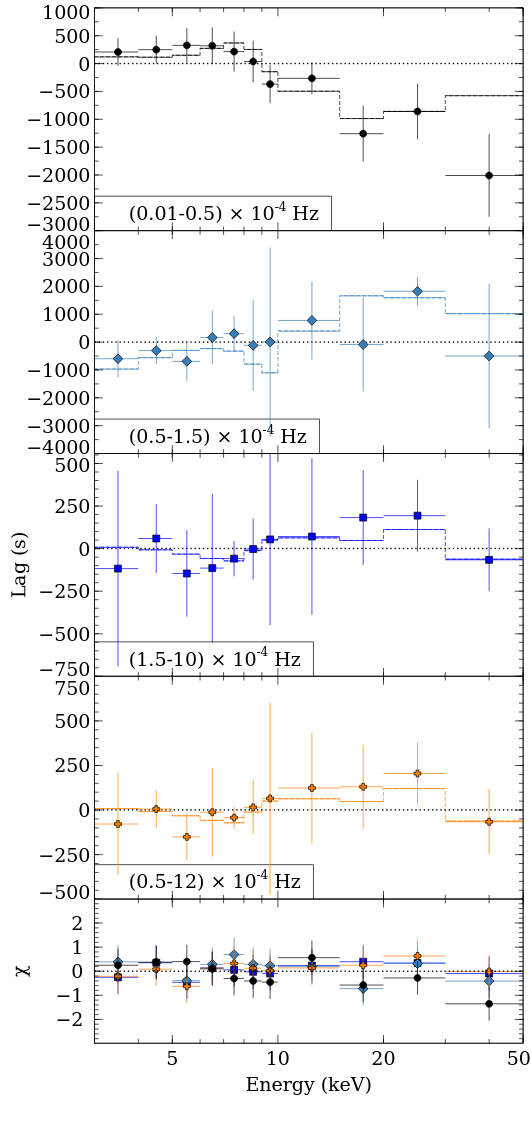}
\caption{Modelled lag spectra of IC\,4329A. Different frequency ranges are shown in different colors and symbols. The spectra are fitted using the relativistic transfer function model {\sc kynreverb}. The dashed lines represent the model lags whereas the error bars represent the observed lags over energy bins.}
\label{modelling}
\end{figure}

\subsection{Modelling lag-energy spectra}

The directly observed continuum from the corona should show intrinsic variability which needs to be correlated with the reflection-dominated emission from the accretion disc to allow the reverberation lag detection to be statistically significant.
Although the lag spectra provide a time delay between the variability of X-ray emissions from the corona and disc, in practice both emissions contribute a fraction to each other. This causes the lag dilution effect, where the absolute value of the measured lag is diminished \citep[see also][]{Uttley2014}. With current detectors, it is difficult to obtain high-cadence light curves to separate out the components of the direct continuum and the reflected emission, while it is generally the case for optical reverberation mapping.
Even in Fourier space in X-ray/UV/Optical continuum reverberation mapping it has been shown that by performing reverberation modeling one can decouple the components responsible for reprocessing and the BLR continuum \citep[e.g.][]{Cackett2022, Lewin2023}.

X-ray reverberation modeling, however, emerged as a powerful approach that has been performed in Fourier space to separate out the disc and coronal components from the lag spectra. Here, we use the transfer function model {\sc kynreverb}\footnote{https://projects.asu.cas.cz/stronggravity/kynreverb} \citep{Caballero2018} to fit the observed lag spectra. The model is based on the full treatment of general relativistic ray-tracing calculations in vacuum, allowing one to measure the path-length difference between the paths from the corona to the observer and the accretion disc to the observed. Reprocessed emissions from each radius of an ionized disc are calculated self-consistently from the reflection table {\tt xillver} \citep{Garcia2013, Dauser2014} or {\tt reflionx} \citep{Ross2005}. The dilution effect is encoded in the model, which uses the directly observed continuum from the corona and the reflected X-ray flux from the accretion disc.

We fit the four lag-energy spectra simultaneously in {\sc xspec} (v.12.14.1) \citep{Arnaud1996}, which are obtained for different frequency ranges. The best-fit lag spectra are shown in Figure~\ref{modelling}. During the fit, we set some parameters to the values obtained from the time-averaged flux spectrum similar to the fit of \citet{Alston2020}. We fix the iron abundance $A_{\rm fe}$ from the fits of the time-averaged spectrum (see the next section). The 2--10 keV luminosity of the source is set to $L_{2-10~\rm keV}=6.03\times10^{43}~\rm erg~s^{-1}$ obtained by \citet{Brenneman2014} -- the value is roughly the same if we estimate from the {\it NuSTAR} spectrum by extrapolating to 2 keV. The accretion rate measured for this source was 0.21 times the Eddington unit \citep{Markowitz2009}, which we use here. 
Additionally, we set the outer radius ($R_{\rm out}$) at $400~R_{\rm g}$. The black hole mass ($M_{\rm BH}$), coronal height ($h$), black hole spin ($a$), inner radius ($R_{\rm in}$), photon index ($\Gamma$), and disc inclination ($i$) are left free to vary independently but are tied between the spectra. These parameters with errors at 90\% confidence level have been found to be at $M_{\rm BH} = 1.37_{-0.36}^{+0.33}\times10^8~M_{\odot}$, $h = 2.45_{-2.36}^{+1.92}~R_{\rm g}$, $a =0.97_{-0.26}^{+0.02}\,{GM}/c^2$, $R_{\rm in}=1.32_{-0.26}^{+34.16}\,R_{\rm g}$ ($1.32_{-0.12}^{+0.19}\,R_{\rm g}$ with 68\% confidence error), $\Gamma = 1.57_{-0.29}^{+0.23}$ and $i = 87.09_{-13.4}^{+0.81\,\circ}$. From the fit we find the disc electron density of log$(n_{e})=15.14_{-0.12}^{+0.67}~\rm cm^{-3}$, which turns out to be roughly the same value obtained for the X-ray bright NLS1 Ark 564 ($15.1_{-0.1}^{+0.2}~\rm cm^{-3}$; \citet{Lewin2022}). For both sources, it has been noticed that if the densities take a higher value, the fit yield becomes poor, while these values from the spectral analysis have been found at $<19.5~\rm cm^{-3}$ in this work for IC\,4329A and $18.55~\rm cm^{-3}$ for Ark 564 \citep{Jiang2019}.


The best-fit spectra provide a reduced $\chi^2_{\rm \nu}$/d.o.f. = 16/30.
We estimate the errors from the Markov Chain Monte Carlo (MCMC) analysis. For that we used the Jermy Sandars {\sc xspec\_emcee}\footnote{https://github.com/jeremysanders/xspec\_emcee} script in {\sc xspec}, which is a pure Python based the Goodman \& Weare's Affine Invariant MCMC Ensemble sampler. Here, we generate 50 MCMC chains with length 10,000 and burn the first 1000. 

\subsection{Modelling time-averaged flux spectra}

We turn now to the spectral analysis to account for the reflection hump and associated relativistic effect, which is imprinted on the X-ray spectrum. Generally, the hump feature in the lag-energy spectrum is likely to appear with a relative time delay, while it makes the flux spectrum be relativistically blurred. With this motivation, we started fitting the time-averaged spectrum using the relativistic reflection model {\sc relxillCp}, a variant of the {\sc relxill} package \citep{Garcia2014, Dauser2014} -- which describes the X-ray reflection spectrum being due to irradiation of the disc by a broken power law emissivity. The model convolves the reflection code {\sc xillver} \citep{Garcia2013} and the ray-tracing code {\sc relline} \citep{Dauser2013} to account for the relativistic blurring effect seen in the soft excess, broadened Fe K line, as well as the reflection hump. 
The use of a particular component {\sc relxillCp} is that it can describe the incident X-ray spectrum by the thermal Comptonization continuum with {\sc nthComp} \citep{Zdziarski1996, Zycki1999}, which has the free parameter the high-energy cutoff -- parameterized by the electron temperature of the corona $kT_{\rm e}$. 

\citet{Brenneman2014} provided an estimate of the coronal temperature of IC\,4329A from the combined fits of the {\it Suzaku} and {\it NuSTAR} spectra obtained from 2012 observations. Using the Comptonization model {\tt compTT} \citep{Titarchuk1994} in combination with non-relativistic reflection model {\sc xillver} the authors derived a coronal temperature of $kT_{\rm e}=50_{-3}^{+6}$ keV for the spherical geometry. Additionally, the modelling includes the {\tt zgauss} component to fit the Fe K line and an ionized absorber with the {\sc xstar} model to fit the X-ray band down to 0.7 keV of {\it Suzaku}.

Here, we explore the relativistic reflection modelling on the 3--50 keV X-ray spectrum (the energy range is consistent with the lag-energy spectra) produced from the long {\it NuSTAR} observation. We present that a single {\it NuSTAR} spectrum can be used to constrain the high-energy cutoff from reflection spectroscopy -- fitting the reflection hump with {\sc relxill} -- the internal resolution of the model is set up to fully relies on the {\it NuSTAR} instrumental responses. Consequently, we specifically rebin the {\it NuSTAR} data to ensure appropriate sampling, to reshape the spectral resolution at the entire X-ray band, whereas the {\it NuSTAR} resolution is 0.4/0.9 keV at 6/60 keV. 
\citet{Garcia2015} demonstrated within the framework of reflection modelling, where {\it NuSTAR} data that have a default uniform {\it NuSTAR} bin width of 0.04 keV throughout the energy band were further binned to 0.2 keV for below 20 keV and 0.4 keV for above 20 keV, respectively. 
The authors noted that without this binning, the fitted values of the high-energy cutoff parameter ($E_{\rm cut}$) were consistently biased below the input. By resampling the {\it NuSTAR} data in reflection spectroscopy, \citet{Kara2017} placed a significant constraint on the coronal temperature with the lowest value of $\sim15$ keV in Ark 564. In \citet{Barua2020}, we did a follow-up by rebinning the same {\it NuSTAR} data, which allowed us to observe a variation in temperature.
Similarly, we resample the {\it NuSTAR} data of IC\,4329A from its default bin width to 0.2/0.4 keV for above and below 10 keV. 

\begin{figure}
\vspace{-2.0cm}
\hspace{-1.7cm}
\includegraphics[scale=0.45, angle=-90]{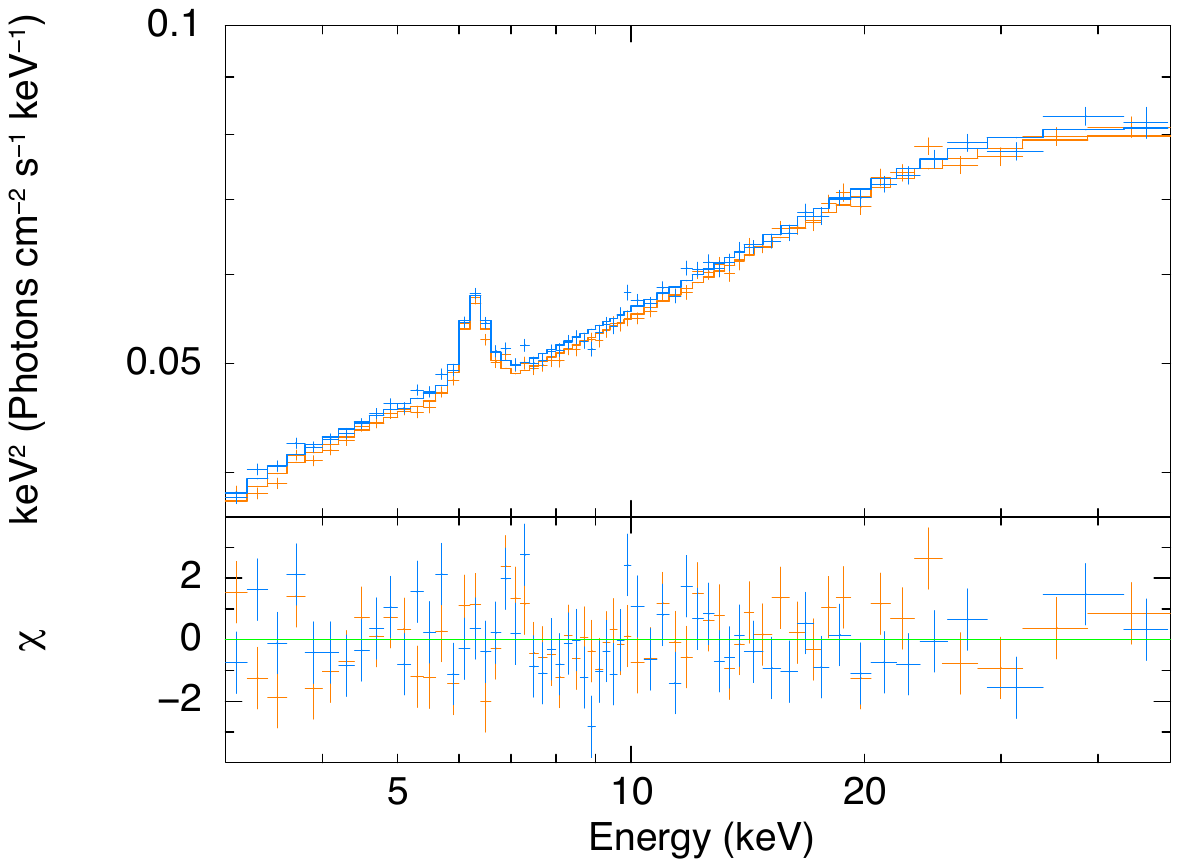}
\caption{{\it NuSTAR} spectra of IC\,4329A fitted using relativistic reflection model {\sc relxillCp}. Shown are the binned spectra from FPMA(blue) and FPMB (orange) fitted simultaneously. The default {\it NuSTAR} bin width of 0.04 keV is changed to 0.2/0.4 keV above and below 10 keV (see text).}
\label{flux_spectra}
\end{figure}

\begin{table}
\centering
\caption{Results of the spectral fit to the time-averaged spectra from {\it NuSTAR} using the relativistic reflection model. The electron density of the accretion disc and the black hole spin take values of $>19.5~\rm cm^{-3}$ and $>0.98$. The {\bf galactic absorption} is fixed at $N_{\mathrm{H}}$ = $4.61 \times 10^{20}$ cm$^{-2}$. The complete model is a composite of {\sc Tbabs $\times$ zTbabs $\times$ (zgauss+relxillCp)}.}
\setlength{\tabcolsep}{1.4pt}
\begin{tabular}{llllllll}
\hline
Model & Parameters &  Values  \\\\
\hline
\hline
{\sc zTbabs} &  $N_{\mathrm{H}} \times 10^{22}$ (cm$^{-2}$) & $1.17^{+0.22}_{-0.19}$ & \\\\

{\sc zgauss} & Line Energy / E (keV)  & $6.36^{+0.03}_{-0.03}$ \\\\

  &Line Width / $\sigma$ (keV)  & $0.21^{+0.06}_{-0.05}$ \\\\

  & Normalization / $A_{\rm gauss}\times10^{-4}$  & $1.24^{+0.17}_{-0.16}$\\\\

{\sc relxillCp} & Photon index  / $\Gamma$ & $<2.40$ \\\\

            & Ionization /  log\,$\xi ~(\rm erg~cm^{-2}~s^{-1})$  & $4.36^{+0.02}_{-0.03}$ \\\\

            & Iron Abundance /  $A_{\rm fe}$  & $3.01^{+0.02}_{-0.03}$\\\\

            & Coronal Temperature / $kT_{\rm e}$ (keV)  & $50.26^{+5.58}_{-4.03}$\\\\

            & Reflection Fraction / $R$  & $1.52^{+0.09}_{-0.10}$\\\\


            & Inclination / $i$ (degree)  & $77.26^{+0.57}_{-0.67}$\\\\

            & Normalization /  $A_{\rm refl} \times 10^{-4}$ & $1.71^{+0.15}_{-0.16}$ \\\\


&Reduced $\chi^2/{\mathrm{d.o.f.}}$ & 282/257  \\
\hline
\hline
\end{tabular}
\label{table1}
\end{table}

We fit here the rebinned version of the spectra with the reflection model with the inclusion of the {\sc zgauss} component to account for the Fe K line at $\sim6.4$ keV. The fit also requires intrinsic absorption, which has been described by {\sc zTbabs}, while we use {\sc Tbabs} for fixed galactic absorption with a column density of $N_{\rm H} = 4.61\times10^{20}~\rm cm^{-2}$ \citep{Kalberla2005}. Our purpose of this fit with the {\it NuSTAR} spectra is to test reflection spectroscopy and determine the high-energy cutoff via the reflection hump. This has been well constrained here, with the coronal temperature of $50.26_{-4.03}^{+5.58}$ keV, which is consistent with the previous finding of the joint {\it Suzaku} and {\it NuSTAR} fit performed by \citet{Brenneman2014}. We further note that the reflection fraction (ratio of the reflected flux to the continuum flux), which is self-consistently constrained by the {\sc relxill} model, is significantly high, with a value of $R=1.5_{-0.10}^{+0.09}$, favoring the reflection dependency of the source. During the fit, we fix the inner radius of the disc at $1~R_{\rm g}$ and the outer radius at $400~R_{\rm g}$, while the black hole spin has been obtained to be at the maximum allowed value of $>0.98$. The best-fit parameters are shown in Table~\ref{table1}. The used binning of 0.2/0.4 keV below/above 10 keV is chosen just for visual clarity. If we choose instead the break point at 20 keV, no noticeable difference can be seen in the $kT_{\rm e}$, most likely due to the hump feature appearing at $>20$ keV in the spectrum.  We show the best-fit binned spectra from reflection modelling in Figure~\ref{flux_spectra}. The fit was performed in {\sc xspec} and errors were estimated using the {\sc xspec\_emcee} code similar to the fit of the lag spectra. 

\section{Discussion}

We have observed reflection hump reverberation in the X-ray bright Narrow-line Seyfert\,1 galaxy IC\,4329A. 
The energy-dependent lags show a delayed response of the 20--30 keV band (where the reflection hump peaks), with a lag amplitude of $\sim1825$ s at Fourier frequencies $(0.5-1.5)\times10^{-4}$ Hz. The lag amplitude drops to $\sim195$ s as the frequencies increase to $(1.5-10)\times10^{-4}$ Hz. While evidence of the reflection hump was found in the lag spectra in previous works, IC 4329A provides the strongest evidence to date for reverberation associated with the Compton hump.

We carry out general relativistic transfer function modelling, which provides a good description of the observed lags beyond 10 keV X-ray, suggesting a relativistic reverberation origin of the hump feature. The lag spectra calculated for four different sets of frequencies are fitted simultaneously. Our reverberation modelling has placed an independent timing-based measurement of the black hole mass of $M_{\rm BH} = 1.37_{-0.36}^{+0.33}\times10^8~M_{\odot}$ and the coronal height of $h = 2.45_{-2.36}^{+1.92}~R_{\rm g}$. Within the uncertainties, the black hole mass is consistent with the previous measurement of $M_{\rm BH}=1.2\times10^8~M_\odot$ \citep{deLaCalle2010}. 

\begin{figure}
\centering
\includegraphics[scale=0.2, angle=0]{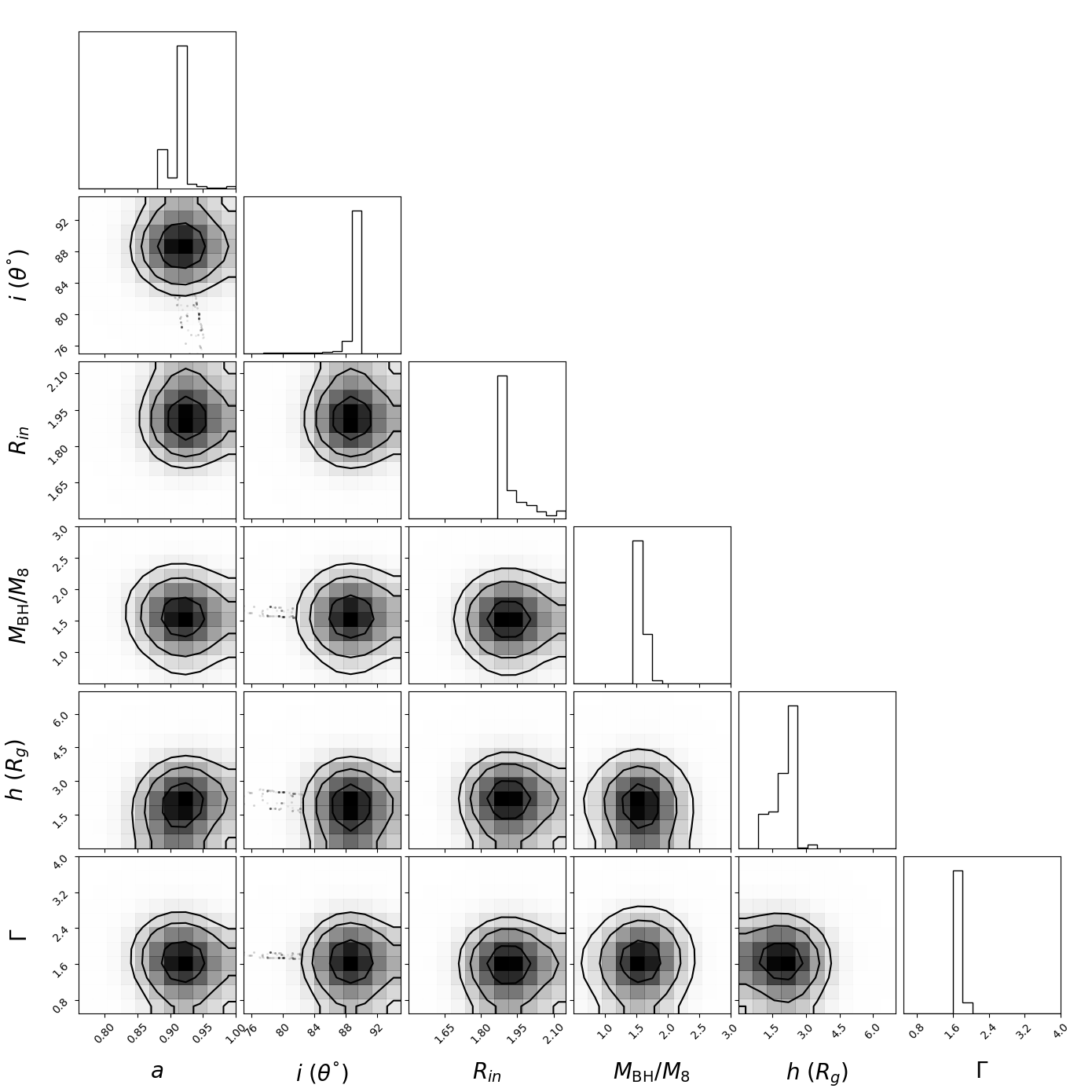}
\caption{MCMC corners from the fit of the lag spectra. The one-dimensional histograms represent the posterior probability distribution, which are normalized to have a total area of one. Shown are the black hole spin $a$, inclination $i$ ($\theta$), inner radius $R_{\rm in}$ ($R_{\rm g}$), black hole mass $M_{\rm BH}$ ($\times10^8\,M_\odot$), coronal height $h$ ($R_{\rm g}$), and photon index $\Gamma$.}
\label{corner_plot}
\end{figure}

It is worth stressing that there appears an inherent degeneracy between the black hole mass $M_{\rm BH}$ and the coronal height $h$ as well as the other parameters, inclination $i$, inner radius $R_{\rm in}$, spin $a$, and photon index $\Gamma$ (shown in Figure\,\ref{corner_plot}). The model degeneracy between the mass and the coronal height has remained a long standing issue while using a single epoch observation, yielding a static picture \citep{Emmanoulopoulos2011, Chainakun2016, Wilkins2016, Chainakun2019, Ingram2019}; Roy et al.\,({\it in preparation}). However, by performing simultaneous modelling over 16 long {\it XMM-Newton} observations \citet{Alston2020} was able to overcome this, showing that multiple observations could help break the degeneracies. While degeneracies are notable in our analysis, with the posterior distributions appearing not to be perfectly Gaussian, as shown in \citet{Alston2020} and Barua et al.\,({\it in preparation}), it could inhibit placing a significant constraint on the key physical parameters. Furthermore, we find inconsistency in the electron density of the accretion disc between the fits of the lag and the flux spectra. The observed degeneracy combined with the unreliable measurement of the disc density perhaps leads to a higher disc inclination than it should typically be (fixed at $60$ degrees in \citet{Brenneman2014}).

The presented lag measurements have been achieved currently with {\it NuSTAR} due to its high sensitivity far beyond 10 keV, allowing detection of the hump-like feature. 
While the reflection hump also appears notably in the flux spectra, fitting this feature in the X-ray spectrum using the relativistic reflection model provides constraints on the high-energy cutoff. Reflection spectroscopy enables the measurement of the coronal temperature at $50.26_{-4.03}^{+5.58}$ keV. The shape of the hard X-ray turnover beyond 10 keV was found to be sensitive to the coronal temperature, where we observed that a lower temperature causes more curvature \citep[see Figure\,2;][]{Barua2021} -- and a clear hump-like feature is apparent. The effect of low temperature has also appeared in the lag spectrum. \cite{Lewin2022} suggested that the low temperature corona of Ark 564 could be responsible for an atypical change in the slope of hard lags. In the observed high frequency lag-energy spectra of IC\,4329A, the hump-line feature is apparent with a relative time delay in the 20--30 keV band. This is consistent with its measured low temperature corona obtained from the analysis of the reflection hump. Both the lag and flux spectra undergo relativistic blurring as the modelling takes into account the underlying effects, providing black hole spin to be at maximum allowed limit. For further demonstration of the relativistic blurring effect in IC\,4329A we produce Figure~\,\ref{model_blurring}  using the same reflection model {\sc relxillCp} for two spin cases; $a=0$ and $a=0.99$ \citep[see also][]{Reynolds2021}. 

Looking for relativistic reverberation of the reflection hump has been completed previously in four AGNs, but robust lag detections seems to remain a challenge. Indication of the reverberating response was found in SWIFT J2127.4+5654 and NGC 1365 in their lag-energy spectra \citep{Kara2015NGC1365}. Subsequently, tripling the source \citet{Zoghbi2021} found no evidence for relativistic reverberation in MCG-5-23-16, while the lag spectra of SWIFT J2127.4+5654 were found to show inconsistency with a static lamppost model. 
Despite an unclear peak in the 20--30 keV band in Ark 564, which is the fourth target explored for beyond 10 keV reverberation lags, its lag spectra indicated the presence of the so-called reflection hump in an intermediate frequency range \citep{Lewin2022}. The observed energy-dependent lags can be described by the relativistic reverberation model.
However, fortunately, the lag spectra of IC\,4329A are observed to pick out the 20--30 keV lag, which have been modelled using the lamppost geometry assumed in the {\sc kynreverb} code. Until now, including IC\,4329A reverberation model has been tested beyond 10 keV on four targets. It is worth mentioning that {\sc kynreverb} is not the only code to be used for reverberation modelling, one can instead use the {\sc reltrans} code of \citet{Ingram2019} to fit the Fourier lag spectra \citep[see for e.g.][]{DeMarco2021, Mastroserio2021, O'Neill2025}.

Modelling of the reverberating lines emerging from the close vicinity of the black hole, mostly performed over the {\it XMM-Newton} band, 0.3--10 keV, has been shown to be an effective approach to obtain a direct measure of the black hole mass and the coronal height \citep[e.g.][]{Emmanoulopoulos2011, Cackett2014, Chainakun2016, Alston2020}. It has also been shown in \citet{Wilkins2021Natur} for the first time that modelling the light curves, rather than the lag spectra, places a significant constraint on these key physical parameters, which was obtained for the Seyfert 1 galaxy I\,Zw\,1 with a black hole mass of $M_{\rm BH} = 3.1_{-0.5}^{+0.5}\times10^7~M_{\odot}$ and a coronal height of $h = 3.7_{-0.7}^{+1.1}~R_{\rm g}$. Thus far, timing-based heights have been obtained for AGNs with black hole mass range of a few $\times~ 10^6-10^7~M_{\rm BH}$. Ark 564 also fits well in this range with the black hole mass of $M_{\rm BH} = 2.3_{-1.3}^{+2.6}\times10^6~M_{\odot}$ and the coronal height of $h = 9.6_{-0.7}^{+0.7}~R_{\rm g}$ measured from modelling up to 50 keV keV \citep{Lewin2022}. The measured height for this source was found to be typical, given the general scenario that height would be larger for a smaller mass black hole. Our best-fit coronal height of $h = 2.45_{-2.36}^{+1.92}~R_{\rm g}$ is quite reasonable for a larger black hole mass in IC\,4329A -- which falls at the extreme high mass end of the aforementioned range. It has been further noted that a lag of $\sim1825$ s at a frequency of $\sim10^{-4}$ Hz corresponds to a few tens of $R_{\rm g}$, which appears inconsistent with our best fit value of $h$. The underlying cause of the inconsistency appears to be due to the degeneracy with black hole mass (Figure\,\ref{corner_plot}). This is, however, not the only reason; the lag dilution effect, or the transfer function effect, also likely leads to a smaller height in the reverberation modelling.
In future, reverberation modelling across a wide range of black hole masses will provide further insights into the geometrical heights of the corona and hence the arrangement of the accretion disc-corona system in AGN.


\begin{acknowledgments}
We thank the anonymous referee for the constructive comments that have improved the manuscript.
SB acknowledges support from the China Postdoctoral Science Foundation General Fund (grant no. 404985), Shanghai Postdoctoral Excellence Program (grant no. 2024686), Shanghai Postdoctoral Introduction Project, China National Postdoctoral Introduction Project and National Foreign Expert Project, Ministry of Science and Technology (grant no. 20240238). This research has made use of data and software provided by the High Energy Astrophysics Science Archive Research Center (HEASARC), which is an online multi-mission archive facility of the Astrophysics Science Division at NASA/GSFC. The work was derived obtained from observations obtained from NASA {\it XMM-Newton} and {\it NuSTAR}.
\end{acknowledgments}



%
\facilities{NuSTAR \citep{Harrison2013}}

\software{ HEAsoft \citep{HEAsoft2014}}, pyLag \citep{Wilkins2019}, Astropy \citep{2013A&A...558A..33A,2018AJ....156..123A,2022ApJ...935..167A}, Scipy \citep{Virtanen2020SciPy}, Numpy, Matplotlib.

\bibliography{sample701}{}
\bibliographystyle{aasjournalv7}

\appendix


\begin{figure}
\centering
\includegraphics[scale=0.55, angle=0]{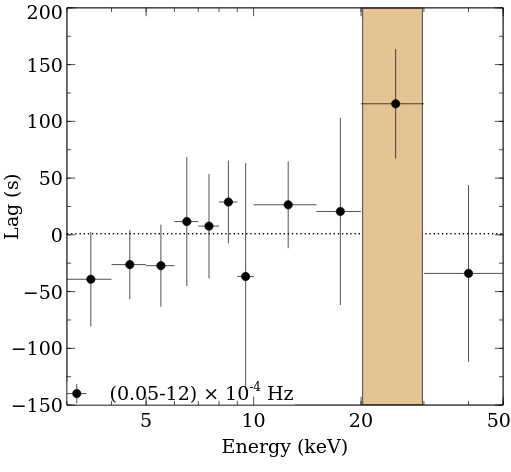}
\caption{Fourier lags versus energy for IC\,4329A computed for frequency range $(0.05-12)\times10^{-4}$ Hz. The delayed response of the 20--30 keV band where reflection hump peaks is clear in the plot (indicated by the shaded region in orange). The lag spectrum picks out the 20--30 keV time lags of $115\pm48$ s.}
\label{low-high-freq_lag_spec}
\end{figure}

\begin{figure*}
\centering
\vspace{-2.0cm}
\hspace{1.0cm}
\includegraphics[scale=0.42, angle=-90]{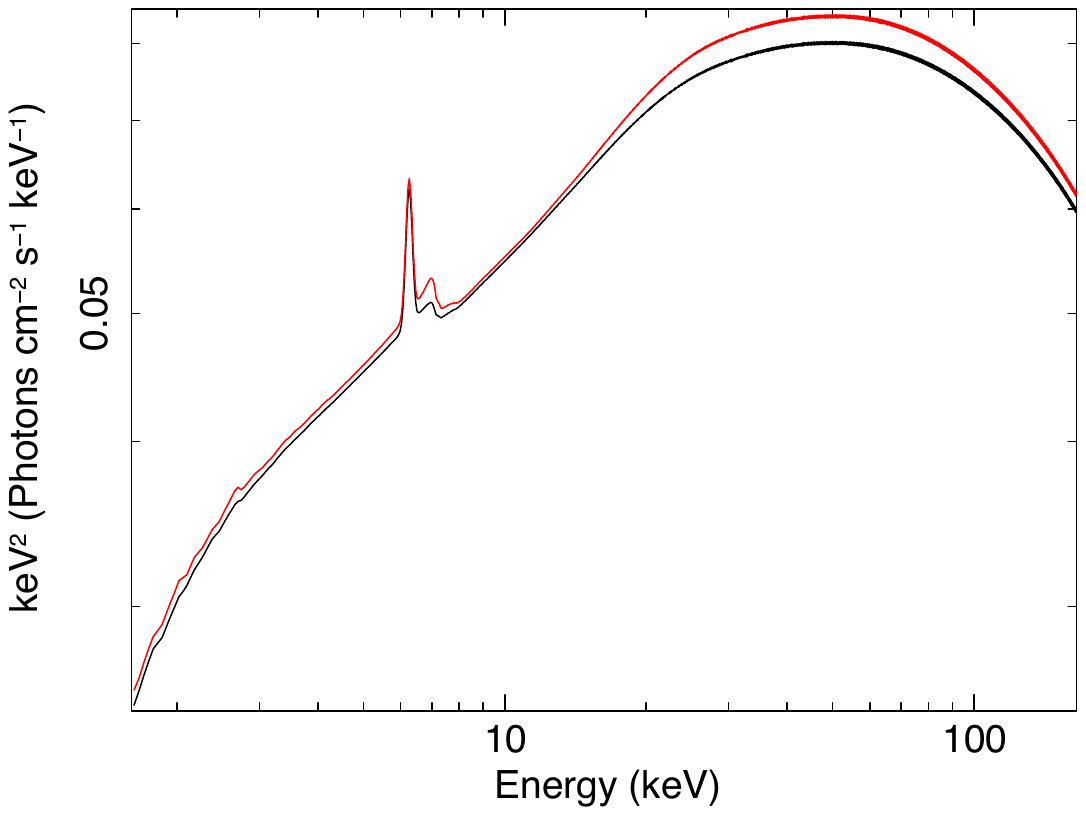}
\vspace{-0.5cm}
\caption{Unfolded modelled spectra of IC\,4329A. The spectra are produced using the {\sc relxill} family model {\sc relxillCp} for two spin cases: spin $a=0$ (shown in red) and $a=0.99$ (shown in black). This plot is a representative of the relativistic blurring effect for the high-spin case.}
\label{model_blurring}
\end{figure*}




\end{document}